\documentclass[twocolumn,twoside,slac_two]{revtex4}
\usepackage{graphicx}
\usepackage{fancyhdr}
\usepackage{graphics}
\usepackage{epstopdf}
\usepackage{textpos}
\usepackage{amsmath}
\usepackage{multirow}
\newcommand{\pt}{\ensuremath{p_{\rm{T}}}}
\newcommand{\et}{\ensuremath{E_{\rm{T}}}}
\newcommand{\abseta}{\ensuremath{|\eta|}}

\begin{document}

\title{\centering Measurement of the inclusive electron cross-section from the decays of heavy flavour hadrons in pp collisions at \boldmath{$\sqrt{s}$}~=~7~TeV at ATLAS}
\author{
\centering
\begin{center}
P. J. Bell on behalf of the ATLAS collaboration
\end{center}}
\affiliation{\centering Universit\'e de Gen\`eve, 1211 Gen\`eve 4, Switzerland}
\begin{abstract}
We present the measurement of the inclusive electron spectrum in proton-proton collisions at a centre-of-mass energy of 7 TeV, using  1.3 pb$^{-1}$ of data collected by the ATLAS experiment at the Large Hadron Collider. Signal electrons in the transverse momentum range  $ 7 < \pt < 26$~GeV and within $|\eta| <2.0$, excluding $1.37<|\eta|<1.52$, are extracted from the dominant hadron and conversion backgrounds. After subtraction of the small  $W$/$Z$/$\gamma^{\ast}$ contribution, the cross-section as a function of~\pt~is found to be in good agreement with theoretical predictions for heavy flavour production from Fixed Order NLO calculations with Next-to-Leading-Log high-\pt~resummation.
\end{abstract}

\maketitle
\thispagestyle{fancy}


\section{INTRODUCTION}

The inclusive production of electrons 
 at low transverse momentum (\pt) is 
dominated by decays of charm and beauty hadrons
and may be used to constrain
theoretical predictions for heavy-flavour production.
In this analysis~\cite{prl} the~\pt~spectra of inclusive electrons
is measured  using an integrated luminosity of~$1.28\pm 0.04$~pb$^{-1}$ of ATLAS data,
within a kinematic acceptance of  $ 7 < \pt < 26$~GeV 
and pseudorapidity{\footnote {ATLAS uses a right-handed 
coordinate system with its origin at the centre of the
detector and the z-axis coinciding with the axis of the beam pipe. 
The pseudorapidity $\eta$ is defined in terms of the polar angle $\theta$ 
as $\eta = - \rm {ln}~\rm{tan}(\theta/2)$.}}
range $|\eta|<2.0$, excluding $1.37<|\eta|<1.52$.

The ATLAS detector~\cite{DetectorPaper} consists of three main components: an  Inner Detector (ID) tracking system, surrounded by electromagnetic (EM) and hadronic calorimeters and an
outer muon spectrometer.
The ID provides precise track reconstruction within
 $|\eta| < 2.5$, employing silicon pixel and microstrip detectors
and an outer Transition Radiation Tracker~(TRT).
Within~$\abseta < 2.5$, EM calorimetry is provided by the 
barrel and end-cap lead/Liquid-Argon (LAr) EM sampling calorimeters.
The reconstruction of electron candidates is seeded by a  preliminary set of clusters in the EM
calorimeter using a sliding window algorithm, with those clusters having a match
to a suitable ID track being reconstructed~\cite{Note}.
In the transition region between the barrel and end-cap calorimeters at $1.37 < |\eta| < 1.52$
the electron identification and energy resolution is degraded by the large amount 
of material in front of the first active layers of the calorimeters, prompting the exclusion of this region from the analysis.


The candidate events were selected using the hardware-based first-level (L1) calorimeter trigger, which identifies EM clusters 
within $|\eta|< 2.5$ above a given energy threshold.
The data were recorded under four different trigger conditions, with
a progressively higher minimum 
cluster transverse energy (\et) requirement applied as the instantaneous luminosity of the LHC increased.
The bulk (76\%) of  the integrated luminosity
used in the analysis was obtained with the L1 calorimeter trigger configured with an 
energy threshold 
of approximately 15~GeV, with the remaining 14\%, 9\% and 1\% recorded with 11, 6 and 3~GeV thresholds, respectively. 

Simulated data samples have been generated in order to estimate backgrounds 
and correct for the trigger and reconstruction efficiencies and 
the resolution of the detector.
{\tt PYTHIA 6.421} with the {\tt MRST LO*}~\cite{mrst2007lomod} PDF set
was used to simulate samples of electrons
from heavy-flavour and $W$/$Z$/$\gamma^{\ast}$  decays, 
and to simulate all sources of background electrons.
All signal and background samples were 
generated at $\sqrt {s} = 7$~TeV using the ATLAS MC09 tune~\cite{mc09Tune}, 
and passed through the simulation of the ATLAS detector.

\section{ELECTRON SIGNAL EXTRACTION}

Events from $pp$ collisions are selected by requiring a collision vertex with more than two associated tracks.
From these events, reconstructed electron candidates are required to
pass a minimum cluster \et~cut
between 7 and 18~GeV depending on the
trigger condition, to lie within the pseudorapidity coverage of the~TRT, $|\eta| < 2.0$, and to be 
outside the transition region between the barrel and end-cap calorimeters.

Preselected candidates must be associated to tracks containing a minimum number of
hits in the ID and are required to pass a minimum
requirement on the fraction of the raw energy deposited in the first layer of the EM calorimeter.
Candidate electrons are then selected from those passing the preselection
by imposing further identification criteria~\cite{Note} designed to suppress electron-like (fake) 
signatures from hadrons. 

\begin{figure}[h]
\begin{center}
\begin{tabular}{cc}
\includegraphics[width=0.48\textwidth]{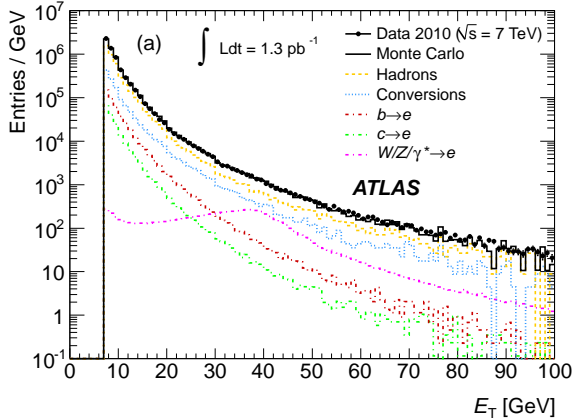} 
\end{tabular}
\end{center}
\caption{Distribution of cluster transverse energy, \et,~for the electron candidates~\cite{prl}.
 The simulation uses {\tt PYTHIA} with the $W$ and $Z/\gamma^{\ast}$ components 
normalised to their NNLO total cross-sections and the heavy-flavour, conversion and hadronic components  
then normalised to the total expectation from the data.
Data with $\pt < 18$~GeV are rescaled to 1.3~pb$^{-1}$ from lower integrated luminosities. 
}
\label{fig:ElectronKinematics}
\end{figure}

 The cluster transverse energy spectrum for the selected electron candidates in the data and simulation is shown 
in Fig.~\ref{fig:ElectronKinematics}. 
The  candidates in the simulation are sub-divided according to their dominant origins,
which for $\et<26$~GeV are non-isolated signal electrons from semi-leptonic decays of charm and beauty hadrons ($\sim$10\%),
a background of secondary electrons dominated by electrons from photon conversions  ($\sim$20\%) and the 
dominant background of misidentified hadronic fakes.
 The fraction of isolated signal electrons from $W$/$Z$/$\gamma^{\ast}$ production is also shown. 

In order to extract the
signal electrons from the selected candidates,
 a binned maximum likelihood method is used, based on the distributions of 
the fraction of high-threshold (transition radiation) TRT hits out of all TRT hits measured on the 
track, $f_{\rm TR}$, the number of hits in the inner-most (B-layer) pixel detectors, $n_{{\rm {BL}}}$ and 
the ratio of the measured energy of the EM cluster to the track momentum, $E/p$.
From simulation, a three-dimensional  probability density function (pdf) in these variables is constructed 
for the signal and conversion components.
For the  hadronic background, the shapes of the three template distributions
are described by additional free parameters and are fitted to the 
data.
The likelihood fit is performed in bins of $\eta$ (on which the discriminating distributions depend) and in bins of~\et~in 
the range 7-26~GeV.


The systematic uncertainty on the number of extracted signal electrons arising from
the differences between the data and simulation in the discriminating variables for the signal and conversion components is estimated to be less than 4\%, evaluated by repeating the signal extraction with the signal and conversion 
templates adjusted within their systematic uncertainties.
The impact of the finite statistics of the simulated samples ($<$~2.5\%) and any possible bias in the method (7.3\%)
were 
studied using pseudo-experiment techniques. 
The uncertainty associated with the electron energy scale (3.5\%) has been assessed by varying the electron candidate cluster 
energy by 1\% for $|\eta|<1.4$ and by 3\% for $|\eta|>1.4$, these systematic effects having been evaluated from 
$Z\rightarrow e^+e^-$ events.
Overall a statistical (systematic) uncertainty on the extracted signal component of approximately 3 (9)~\% is obtained.

\section{DIFFERENTIAL CROSS-SECTION MEASUREMENT}

The measured differential cross-section for electrons originating from heavy flavour production
within the fiducial acceptance is defined by
\begin{equation}
\frac{\Delta\sigma_i}{\Delta p_{{\rm {T}}_i}} =  
\left( \frac{N_{\mathrm{sig}_{\,i}}}{\epsilon_{\mathrm{trig}_i} \cdot \int{\cal L}\mathrm{d}t} -
\sigma^{W/Z/\gamma^*}_{\mathrm{accepted}_i} \right) 
\cdot 
\frac{C_{\mathrm{mig}_i}}{\epsilon_{\mathrm{(rec+PID)}_i}} \cdot
\frac{1}{\Gamma_{\mathrm{bin}_i}},
\label{eq:cross_section}
\end{equation}
where 
$N_{\mathrm{sig}_{\,i}}$ is the number of extracted signal electrons with reconstructed
$p_\mathrm{T}$ in bin $i$ of width ${\Gamma_{{\rm{bin}}_{\,i}}}$,
$\int{\cal L}dt$ is the integrated luminosity, 
$\epsilon_{\mathrm{trig}_i}$ is the trigger efficiency and
$\epsilon_{\mathrm{(rec+PID)}_i}$ the combined reconstruction and identification efficiency.
$C_{\mathrm{mig}_i}$ is the bin migration correction factor, 
defined as the ratio of the number of electrons in bin $i$ of true
\pt~and the number in the same bin of reconstructed \pt.
$\sigma^{W/Z/\gamma^{\ast}}_{{\rm{accepted}}_i}$ is the small and well understood 
accepted cross-section
from $W$/$Z$/$\gamma^{\ast}$ production~\cite{firstWZ} which must be subtracted from the total signal.

The efficiencies of the 3 and 6~GeV threshold L1 triggers are measured using events selected by an alternative, 
very inclusive minimum bias
trigger,
and the 11 and 15~GeV triggers using events recorded by the
6~GeV trigger, which is fully efficient in the \et~region for which
the higher threshold triggers are used.
A systematic uncertainty is estimated by comparing the
measured trigger efficiencies to those expected in the simulation for electrons from heavy-flavour decays.

The overall efficiency and migration correction factor, $\epsilon_{\mathrm{(rec+PID)}_i}/C_{{\rm {mig}}_i}$, 
is determined from {\tt PYTHIA}-simulated samples of heavy-flavour decays to electrons.
The statistical uncertainty is between 0.4 and 3.5\% and
a systematic uncertainty of 5-10\% is estimated  by recalculating
$\epsilon_{\mathrm{(rec+PID)}_i}$ $/C_{{\rm {mig}}_i}$ from simulated samples produced with a 5\%
increase in the amount of material in the ID.
Additionally, the efficiency of the electron 
identification cuts in the simulation is compared with a measurement 
made on data using a tag-and-probe (T\&P) technique.
The probe candidates, which must pass only the preselection
cuts, are taken from a sample of 
events enriched in heavy quark pair production 
where both  heavy hadrons decay semi-leptonically. 
The signal purity remains low after the T\&P selection, 
necessitating a method similar to that described in Section 2 to extract the signal component of the probe candidates before and after applying the identification criteria. 
By comparing the measured identification efficiency of the extracted probe electrons to that expected in simulation as a function of \pt, an uncertainty of 5\% is obtained on the identification efficiency, with a further 7\% systematic uncertainty coming from the T\&P method itself. 

\begin{figure}[h!]
\includegraphics[width=0.5\textwidth]{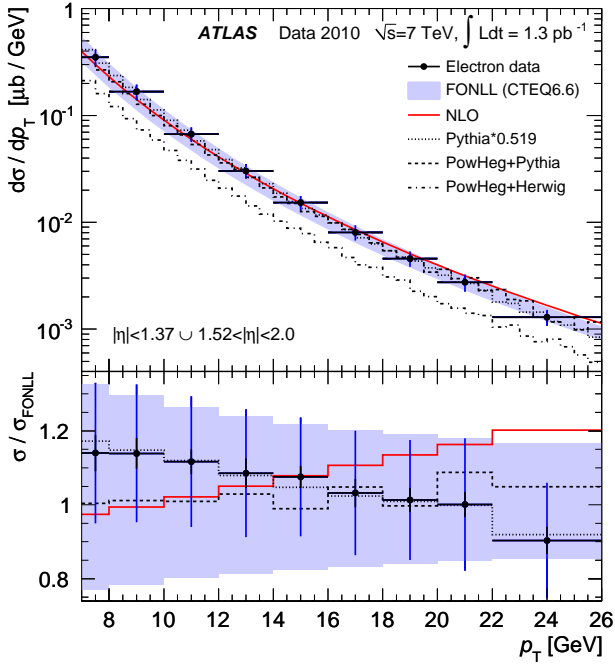}
\caption{Electron differential cross-section as a function of~\pt~for $|\eta| <2.0$  excluding $1.37 < |\eta| <1.52$~\cite{prl}.
The ratio of the measured and predicted cross-sections to the FONLL calculation is given in the bottom.
The {\tt PYTHIA} (LO) cross-sections are normalised to the data.
}
\label{fig:FullComparison}
\end{figure}

The differential cross-section for electrons from heavy-flavour production obtained from Eqn. (1)
is shown in Fig.~\ref{fig:FullComparison}.
A comparison is made to the prediction of the FONLL theoretical framework,
in which the heavy quark production cross-section is calculated in pQCD by matching 
the Fixed Order NLO terms with NLL 
high-p$_\mathrm{T}$ resummation. 
The uncertainties associated with this prediction originate from several different sources. 
The dominant contribution comes from the renormalisation and factorisation scales 
(up to 35\% at low \pt).
The uncertainty on the heavy quark masses contributes up to 9\% at
low \pt, and
the PDF-related uncertainty 
(taken from the  {\tt CTEQ6.6} error set)
is below 8\% over the whole~\pt~ range.
Uncertainties arising from the value of $\alpha_s$ and
on the non-perturbative fragmentation function are  small.
The measured results are seen to be fully compatible with the overall 20-40\% FONLL uncertainty band.

The results are also compared to the NLO predictions of the {\tt POWHEG}~\cite{Powheg1, Powheg3} program, 
interfaced to either {\tt PYTHIA} or {\tt HERWIG}~\cite{herwig} for the parton shower simulation, 
and to the sum of LO and parton shower predictions of {\tt PYTHIA}.
Whereas {\tt POWHEG+PYTHIA} agrees well with the FONLL predictions,  
{\tt POWHEG+HERWIG} predicts a significantly lower total cross-section. 
 {\tt PYTHIA} (LO) describes the \pt-dependence well but predicts approximately a factor two higher total 
cross-section.
The NLO expectation obtained from the FONLL
program by removing NLL resummation from the pQCD calculation shows a clear deviation from the 
data, underlining the importance of the NLL corrections.

\section{CONCLUSIONS}

The differential cross-section of electrons arising from heavy-flavour  production 
has been measured in the electron transverse momentum range 7 $<$ \pt~$ <$ 26 GeV and 
pseudorapidity region  $|\eta|<2.0$ (excluding $1.37<|\eta|<1.52$)
and is found to be in good agreement with the theoretical predictions for heavy-flavour production 
from the FONLL computation.
Good agreement is also seen with the predictions of 
{\tt POWHEG+PYTHIA}, although {\tt POWHEG+HERWIG} predicts a significantly lower total cross-section. 
 {\tt PYTHIA} describes the \pt-dependence well, but overestimates the total cross-section by approximately a factor of two.



\section{Acknowledgments}
We thank M. Cacciari for supplying the predictions from FONLL
and for many useful discussions.


\vskip -2mm
\begin{thebibliography}{9}
\bibitem{prl} ATLAS Collaboration, arXiv:1109:0525 [hep-ex], submitted to Phys. Lett. B (2011)
\bibitem{DetectorPaper} ATLAS Collaboration, JINST, {\bf 3}, S08003 (2008)
\bibitem{Note} ATLAS Collaboration, arXiv:1110:3174 [hep-ex], submitted to Eur. Phys. J. C (2011)
\bibitem{mrst2007lomod} A. Sherstnev and R.S. Thorne, Eur. Phys. J., {\bf C55}, 553-575 (2008)
\bibitem{mc09Tune} ATLAS Collaboration, ATL-PHYS-PUB-2010-002  http://cdsweb.cern.ch/record/1247375 (2010)
\bibitem{firstWZ} ATLAS Collaboration, JHEP, {\bf 12}, 060 (2010)
\bibitem{Powheg1} S. Frixione, P. Nason, C. Oleari, JHEP, {\bf 11}, 070 (2007)
\bibitem{Powheg3} S. Alioli, P. Nason, C. Oleari and E. Re, JHEP, {\bf 6}, 043 (2010)
\bibitem{herwig} G. Corcella, I.G. Knowles, G. Marchesini, S. Moretti, K. Odagiri, P. Richardson, M.H. Seymour and B.R. Webber, JHEP, {\bf 01}, 010 (2001)
\end{thebibliography}
\end{document}